\documentclass{iaus}
\usepackage{graphicx}

\title[Magnetism of the He-weak star HR\,2949] 
      {Magnetism of the He-weak star HR\,2949\footnote{Based on observations
          under ESO programs 073.C-0337, 076.C-0164 and the MiMeS large
          program at the CFHT.}}

\author[Th.\ Rivinius et al.]   
       {Th.\ Rivinius$^{1}$, G.~Wade$^2$, R.H.D.~Townsend$^3$, M.~Shultz$^2$,
         J.~Grunhut$^2$, O.~Stahl$^4$, and the MiMeS Collaboration }

\affiliation{$^1$ESO, Chile; 
$^2$RMC, Canada;
$^3$UW Madison, USA;
$^4$LSW/ZAH Heidelberg, Germany
}

\pubyear{2011}
\volume{272}  
\pagerange{1--2}
\jname{Active OB stars: structure, evolution, mass loss and critical limits}
\editors{C.\ Neiner, G.\ Wade, G.\ Meynet \& G.\ Peters, eds.}
\begin{document}
\maketitle

\begin{abstract}
A magnetic field and rotational line profile variability ({\it lpv}) is found
in the He-weak star HR\,2949. The field measured from metallic lines varies in
a clearly non-sinusoidal way, and shows a phase lag relative to the
morphologically similar He\,{\sc i} equivalent width variations. The surface abundance
patterns are strong and complex, and visible even in the hydrogen lines.

\end{abstract}

\firstsection 
\section{Introduction}
The mid-B star HR 2949 was for many years considered to be non-variable, and
indeed was listed in standard star catalogues.  However, \cite[Rivinius et
  al. (2003)]{riv03} noted that the star is actually variable both
photometrically as well as spectroscopically. They classified it as a He-weak
star.
Here we report the discovery of a magnetic field and describe the line profile
variations ({\it lpv}) due to surface abundance inhomogeneities. The existence
of these inhomogeneities was already proposed by Rivinius et al. (2003), but
could not be investigated further on the basis of solely four spectra.

\section{Observations}
Eight {archival spectra} have been obtained with the {\it FEROS} instrument at
La Silla, providing a resolving power of 48\,000 over a spectral range of 375
to 890\,nm. {Spectropolarimetric (Stokes $V$) data} were taken with {ESPaDOnS}
at the 3.6\,m CFHT in the 2009/2010 season, in total 16 measurements. The
observations were obtained as part of the Magnetism in Massive Stars (MiMeS)
Large Program.

\cite[Koen \& Eyer (2002)]{koe02} quote a {period} of $P=1.9093$\,d from {\it
  HIPPARCOS} photometric data. Re-analysing these data using the Lomb-Scargle
statistic, we obtain $P=1.9083\pm0.00025$\,d.  In the spectra, the helium
lines show the strongest variability.  The time base covered by the spectra is
long, 1999 to 2009, but there is severe aliasing. However, the aliases are
narrow and there is only one peak in reasonable agreement with the Hipparcos
data, at $\boldmath P=1.90871\pm0.00007\unboldmath$\,d.  As the {epoch} we
chose the date of the ESPaDOnS measurement showing the most negative field, so
that our adopted ephemeris is:\\[1mm]
\centerline{$
T_{\rm min B}({\rm HJD}) = 2\,455\,223.987 + 1.90871(7)\times E
$}

\section{Spectral Variations}
The magnetic field as measured by ESPaDOnS (using Least-Squares Deconvolution
applied to metallic lines) confirms the photometric and spectroscopic periods.
Figure~1 shows the respective phased variations. The field varies between
$-760$ and $+200$\,G.
There is clear rotational line profile variability ({\it lpv}), with no line
in the spectrum unaffected. Taking the behavior of the He\,{\sc i} lines as a
starting point, the other lines can be sorted into three variability
morphology groups: {\bf Lines similar to helium} are those of Mg\,{\sc
  i}\,8807, Mg\,{\sc ii}\,4481, O\,{\sc i}\,8446. They show a clear modulation
of EW.  Lines behaving {\bf inversely to He\,{\sc i}}, are Si\,{\sc ii}\,4128
and Si\,{\sc iii}\,4553, Cr\,{\sc ii}\,4588, P\,{\sc ii}\,6043, P\,{\sc
  iii}\,4222, and the Balmer lines. Also these vary in EW with rotation.
The majority of lines show a third pattern, e.g.\ S\,{\sc ii}\,5640, Fe\,{\sc
  ii}\,4949, C\,{\sc ii}\,4267, Ca\,{\sc ii}\,8662, Sr\,{\sc ii}\,4215,
Ti\,{\sc ii}\,4564 and thus shall be called {\bf metal group} behavior. While
these lines show clear {\em lpv}, their EW is rather non-variable.
Finally, some lines show different types of {\bf hybrid} variation patterns.
%

\begin{figure}
\begin{center}
\parbox{\textwidth}{%
\parbox{0.33\textwidth}{\includegraphics[angle=270,width=0.33\textwidth,clip]{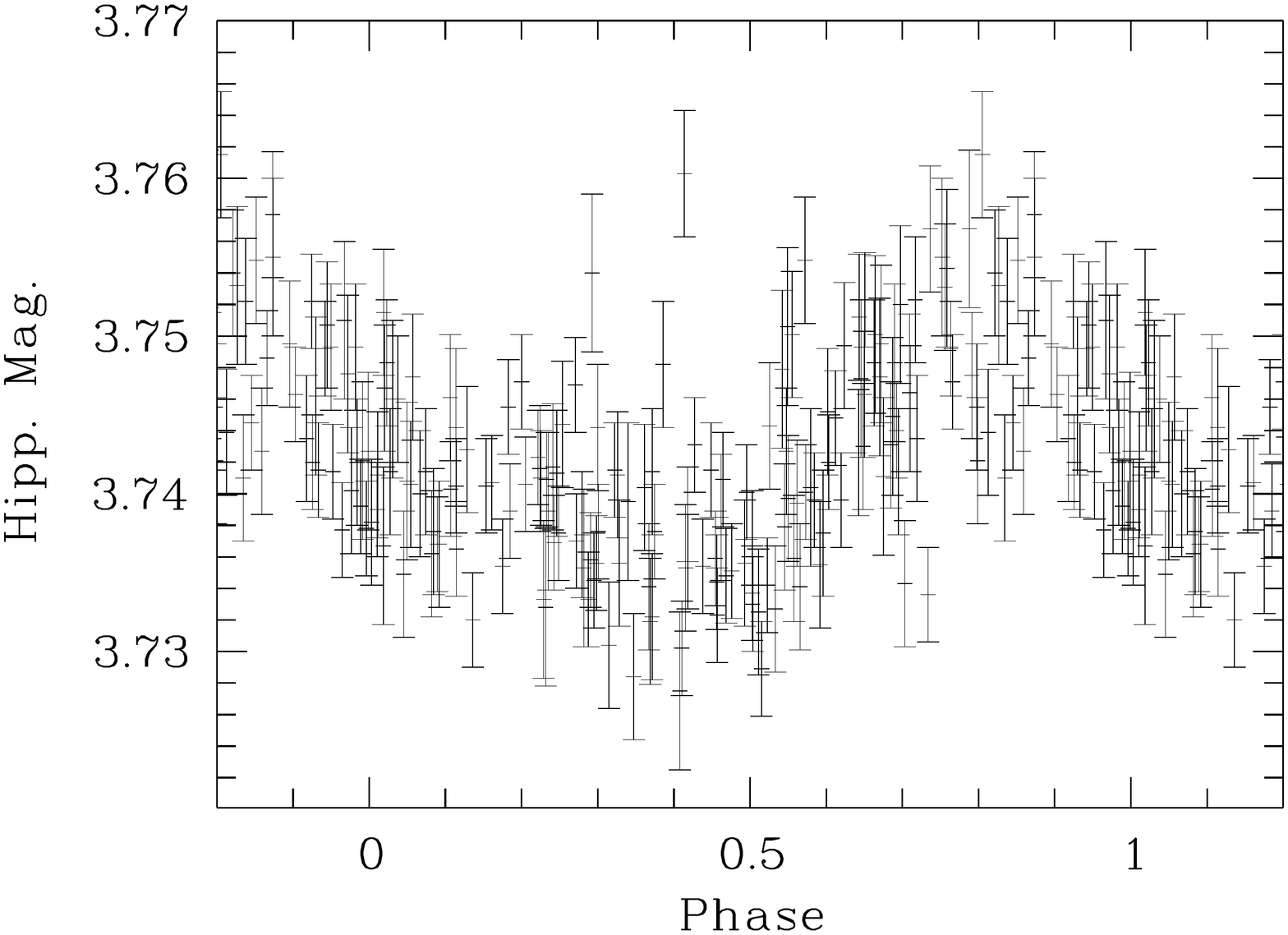}}%
\parbox{0.33\textwidth}{\includegraphics[angle=270,width=0.33\textwidth,clip]{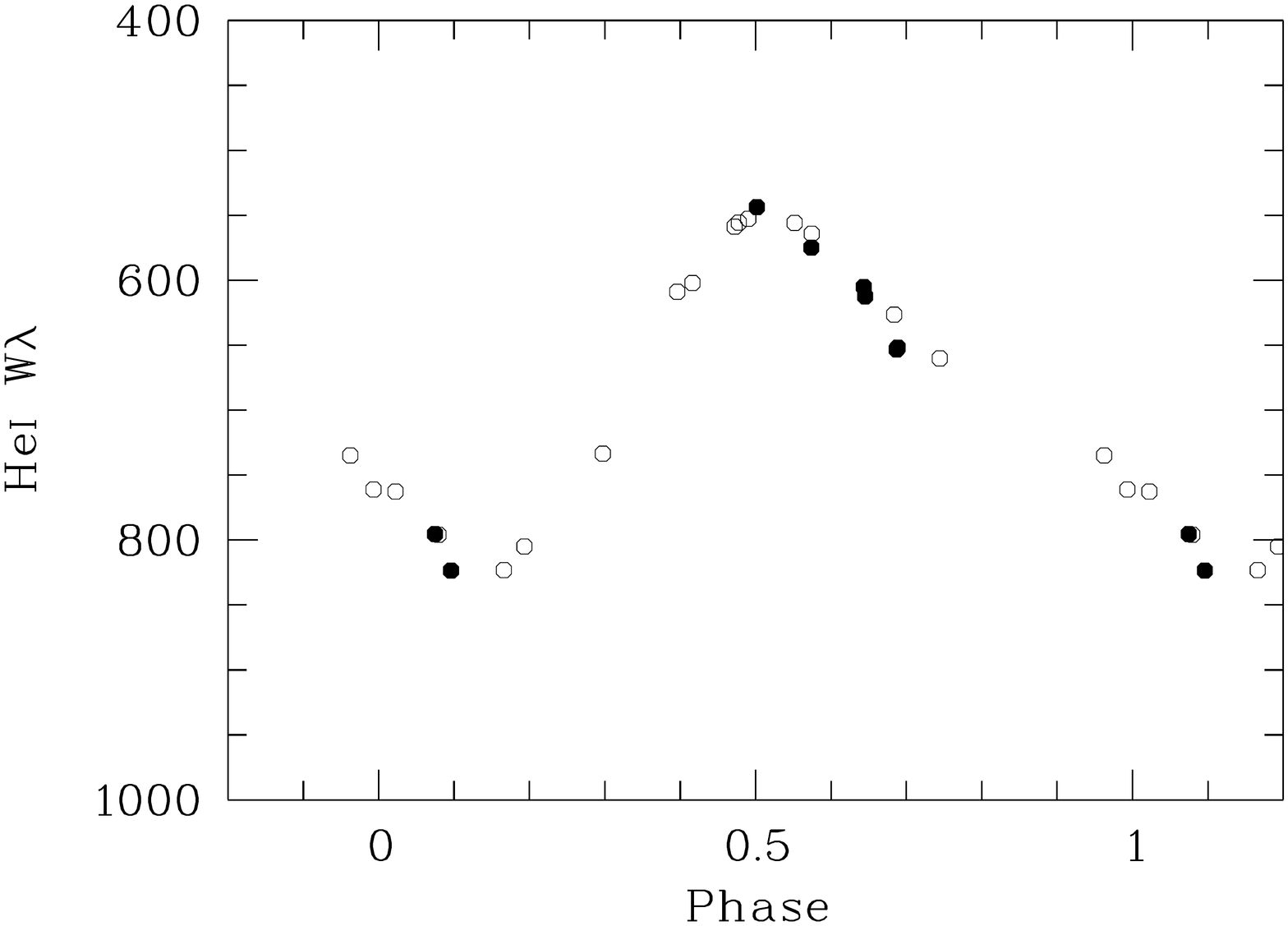}}%
\parbox{0.33\textwidth}{\includegraphics[angle=270,width=0.33\textwidth,clip]{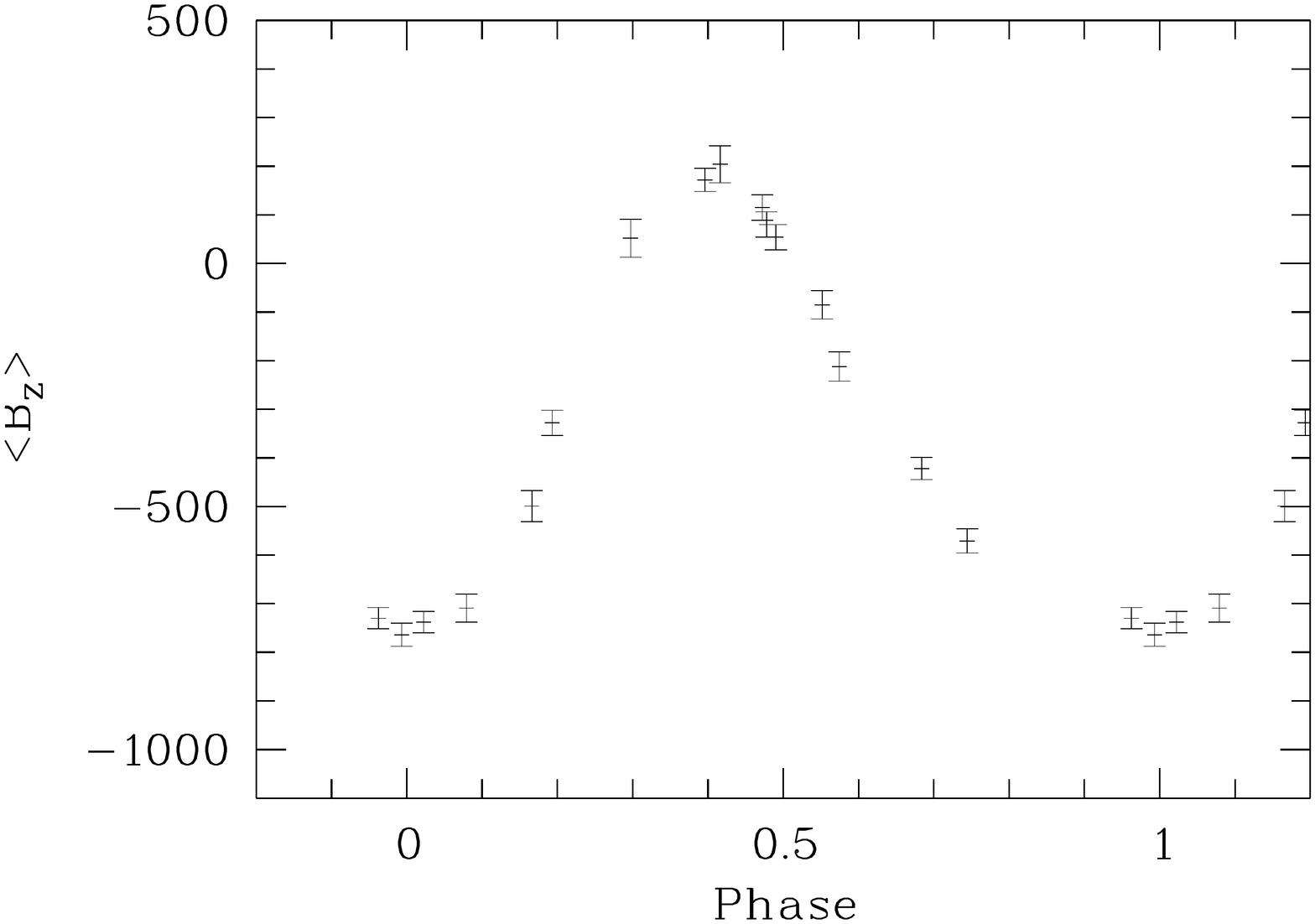}}%
}

\end{center}\begin{center}
\parbox{0.84\textwidth}{
\includegraphics[viewport=70 133 500 750,angle=0,width=0.22\textwidth,clip=]{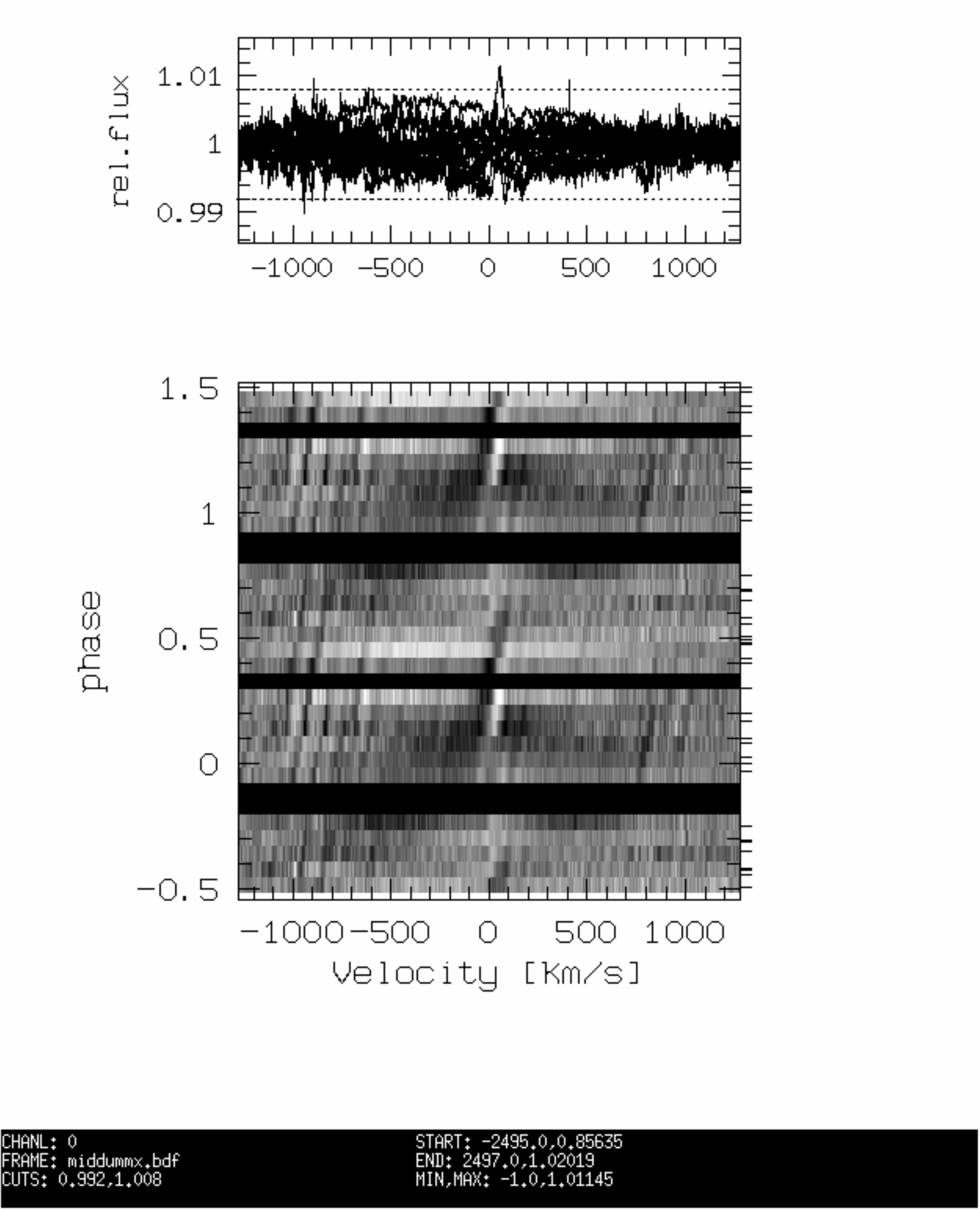}%
\includegraphics[viewport=70 133 500 750,angle=0,width=0.22\textwidth,clip=]{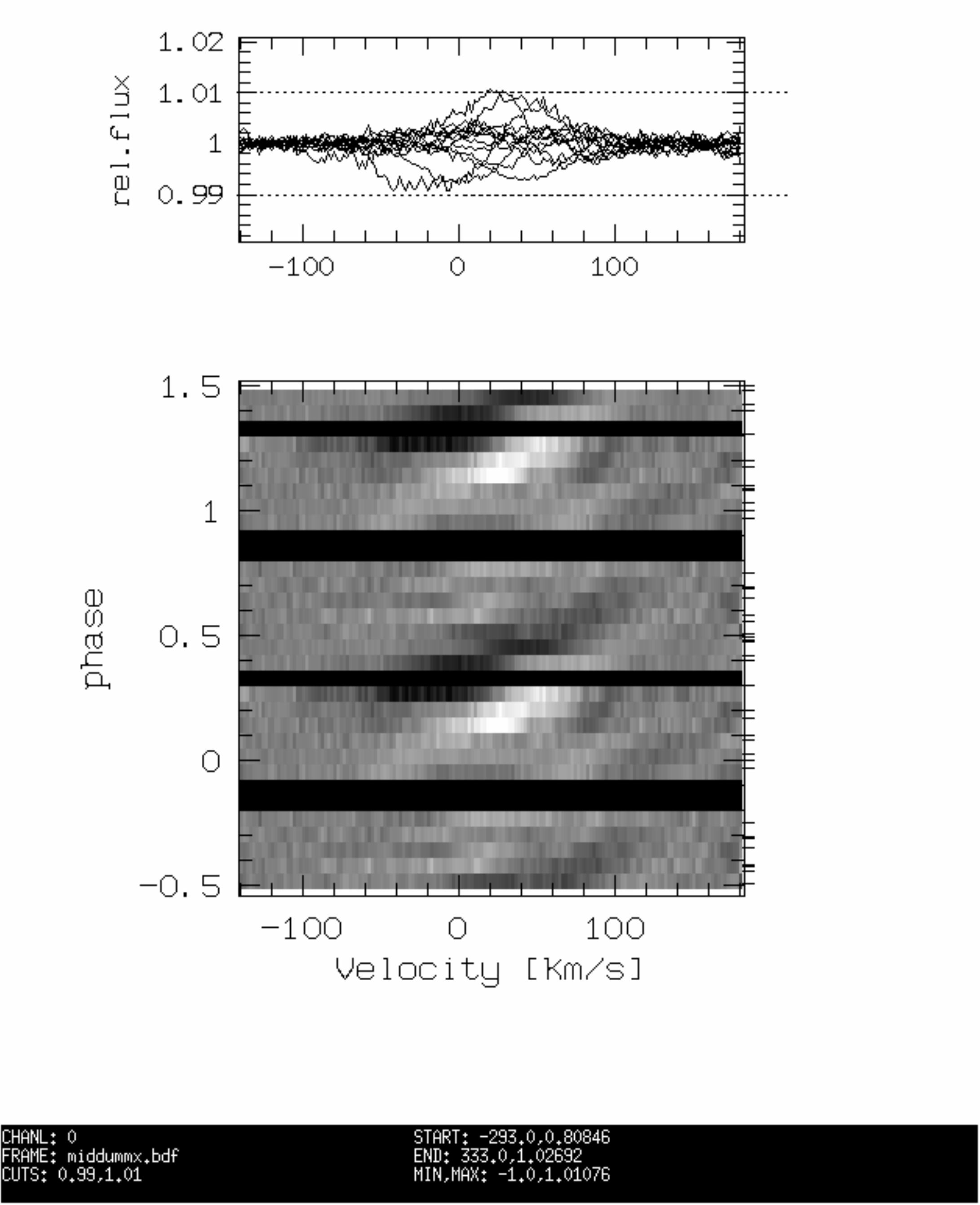}%
\includegraphics[viewport=70 133 500 750,angle=0,width=0.22\textwidth,clip=]{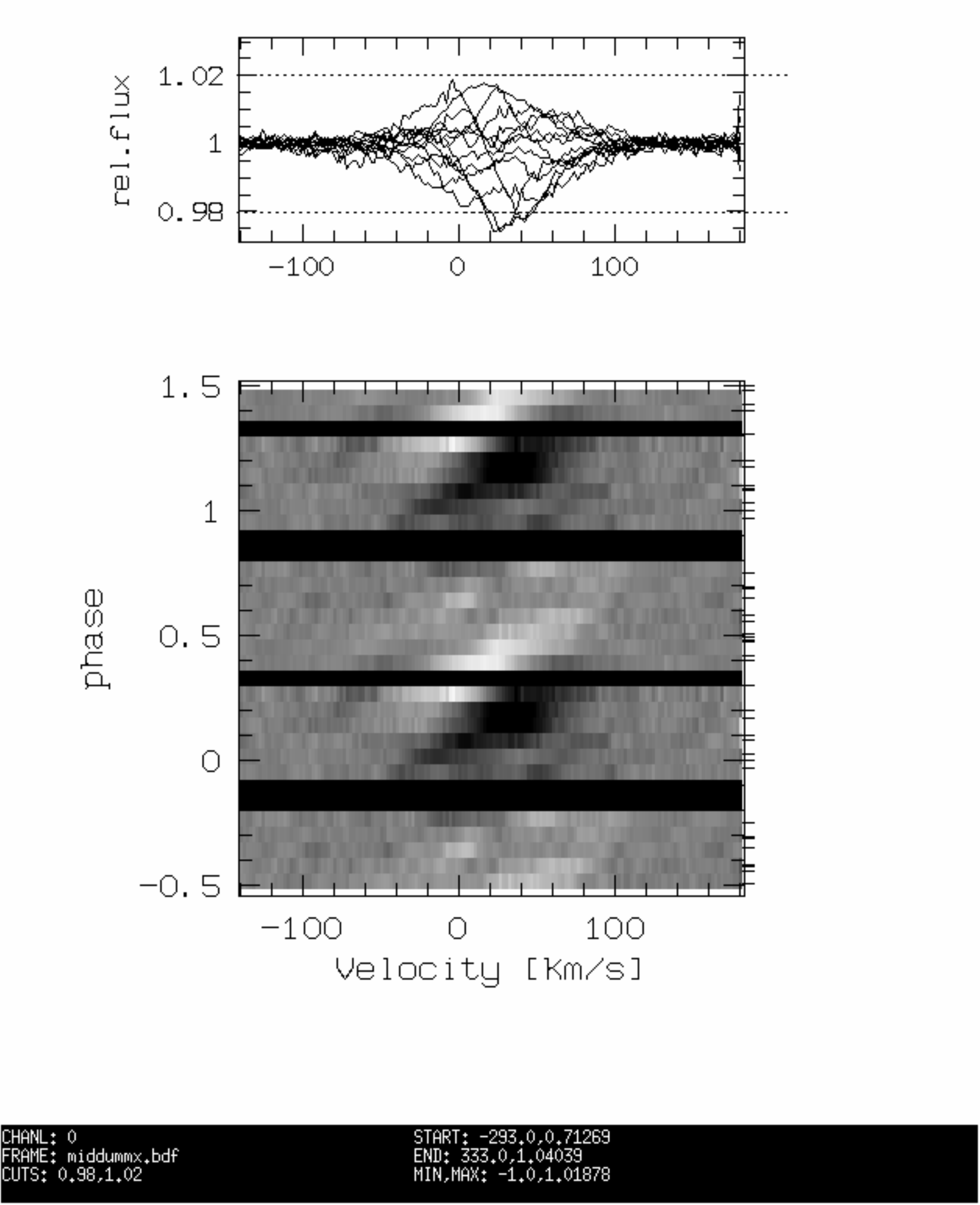}%
\includegraphics[viewport=70 133 500 750,angle=0,width=0.22\textwidth,clip=]{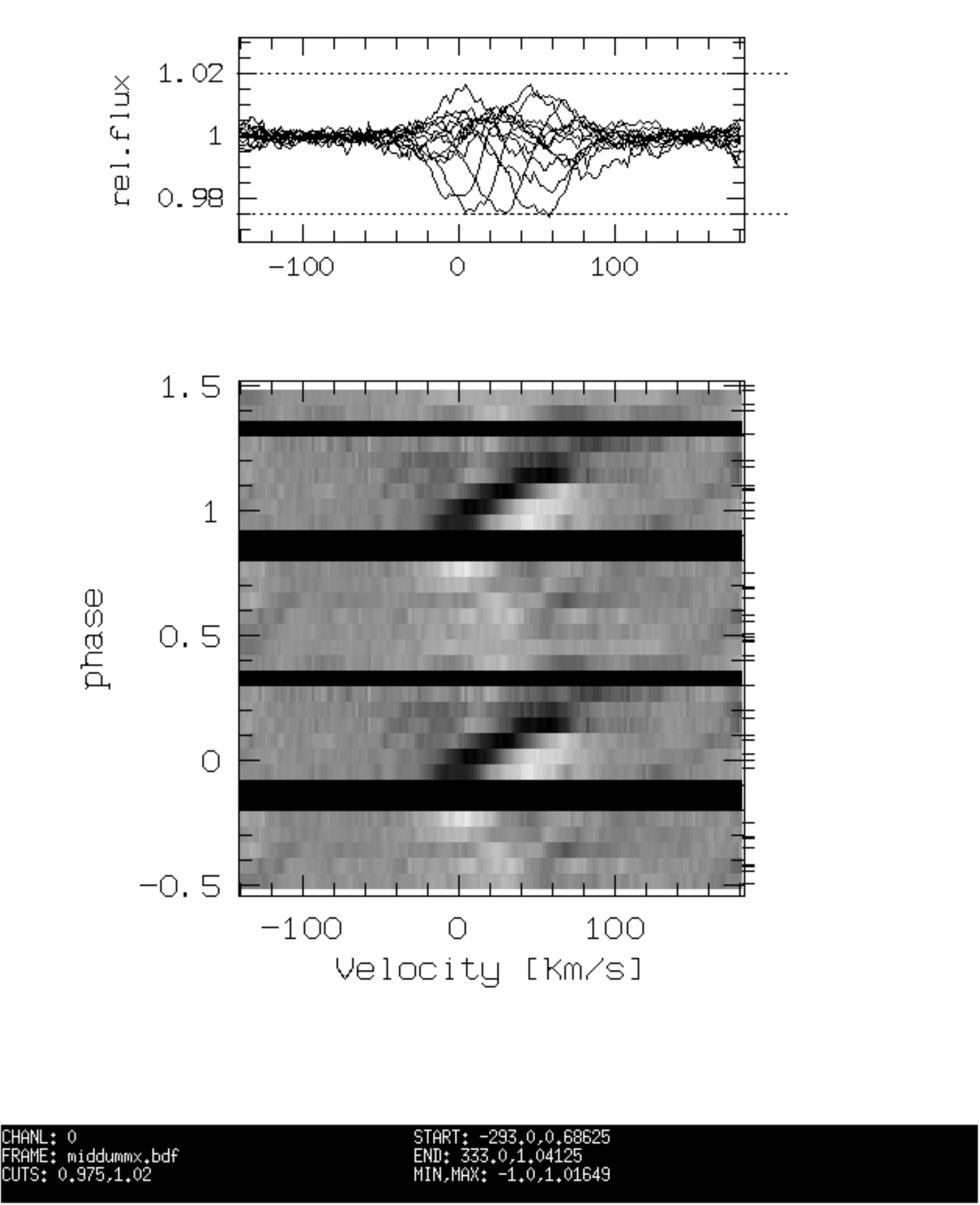}%
}
\end{center}
\caption[]{
\centering
Upper row, from left to right: Hipparcos photometry, equivalent width
variability of the combined He\,{\sc i} 4009 \& 4026 lines (ESPaDOnS, open
symbols; FEROS, filled symbols), and magnetic data.
%
Lower row, left to right: Exemplary types of line profile variation seen in
HR\,2949: Balmer wings, Balmer cores (both H$\beta$), helium (He\,{\sc
  i}\,4388), metal (Fe\,{\sc i}\,4949). }
\label{FNCMa}
\end{figure}

%
%
It is noteworthy that the {Balmer lines} show a clear pattern. The line core
shows an abundance-like pattern well described as an ``anti-helium''-group
behavior. We attribute the pattern to an actual variation of the
{H-abundance}, rather than to a change in local surface parameters.  However,
there are also changes in the Balmer line wings probably caused by {local
  parameter variations}.  Although the effect is subtle, less than 0.5\,\%,
and in spite of the notoriously difficult normalization of the Balmer wings in
echelle data the pattern is identical in all lines.
\section{Conclusions}
HR\,2949 is a {magnetic} star of the {He-weak/Si-strong} type. Its surface
shows the clear signature of the typically complex abundance patterns
characteristic to such stars. In addition to the abundance variation, the
variability of the Balmer line wings indicates some other {local parameter
  changes}, such as pressure or temperature, linked to the magnetic field.

\end{document}